\documentclass[12pt]{article}

\usepackage[dvipdfmx]{graphicx}
\usepackage{amsmath}        
\usepackage{amssymb}        
\usepackage{cite}
\usepackage[dvips]{color}
\usepackage{fancyhdr}

\def\erase#1{{}}
\def\EqArrerase#1{{}}

\makeatletter
 
 \@addtoreset{equation}{section}
\makeatother

\def\GL{{G\kern-.12em L\kern-.04em}}
\def\OSp{{O\kern-.11em S\kern-.04em p}}
\def\IOSp{{I\kern-.06em O\kern-.11em S\kern-.04em p}}
\def\MN{{M\kern-.14em N}}
\def\NM{{N\kern-.14em M}}
\def\NL{{N\kern-.14em L}}
\def\LN{{L\kern-.11em N}}
\def\ML{{M\kern-.14em L}}
\def\LM{{L\kern-.11em M}}
\def\RN{{R\kern-.11em N}}
\def\NR{{N\kern-.14em R}}
\def\RM{{R\kern-.11em M}}
\def\MR{{M\kern-.14em R}}
\def\RL{{R\kern-.11em L}}
\def\LR{{L\kern-.11em R}}
\def\RS{{R\kern-.11em S}}
\def\SR{{S\kern-.11em R}}
\def\SN{{S\kern-.11em N}}
\def\NS{{N\kern-.11em S}}
\def\SM{{S\kern-.11em M}}
\def\MS{{M\kern-.11em S}}
\def\SL{{S\kern-.11em L}}
\def\LS{{L\kern-.11em S}}
\def\sqr#1#2{{\vcenter{\hrule height.#2pt
      \hbox{\vrule width.#2pt height#1pt \kern#1pt
          \vrule width.#2pt}
      \hrule height.#2pt}}}
\def\bra0{\langle0|}
\def\ket0{|0\rangle}
\def\soeji#1_#2#3{#1_{#2}\cdots#1_{#3}}
\def\longgLRarrow{\longleftarrow\kern-3pt\relbar\kern-3pt\relbar\kern-3pt%
\longrightarrow}
\def\longLRarrow{\longleftarrow\kern-3pt\relbar\kern-3pt\longrightarrow}
\def\longLarrow{\longleftarrow\kern-3pt\relbar\kern-3pt\relbar\kern-3pt\relbar}
\def\longRarrow{\relbar\kern-3pt\relbar\kern-3pt\relbar\kern-3pt\longrightarrow}
\def\bothDer#1#2#3{%
\overset{\kern-.7em\stackrel{#1}{#2}}{\partial_{#3}}}
 
\makeatletter
 
 \@addtoreset{equation}{section}
\makeatother

\begin{document}
\thispagestyle{fancy}

\title{Emergence of General Relativity from Cosmological Constant Via Ghost Condensation}

\author{Ichiro Oda
\footnote{Electronic address: ioda@cs.u-ryukyu.ac.jp}
\\
{\it\small
\begin{tabular}{c}
Department of Physics, Faculty of Science, University of the 
           Ryukyus,\\
           Nishihara, Okinawa 903-0213, Japan\\      
\end{tabular}
}
}
\date{}

\maketitle

\thispagestyle{fancy}

\begin{abstract}

We show that starting with a cosmological constant in a curved space-time, the Einstein-Hilbert term 
of general relativity is generated through a ghost condensation. We fix Weyl symmetry, or equivalently 
local scale symmetry by a gauge condition $R = 0$ \`{a} la BRST formalism, and see that the condensation
of the Faddeev-Popov ghosts, $\langle \bar c c \rangle \neq 0$ leads to a generation of the Einstein-Hilbert 
action of general relativity. This dynamical mechanism of symmetry breakdown for a global scale symmetry
is new in the sense that the reduction of fermionic degrees of freedom effectively leads to a generation of 
bosonic degrees of freedom. We also discuss this mechanism from the viewpoint of the problem of a bound state, 
and show that asymptotic fields corresponding to the bound states are ``confined'' to the unphysical Hilbert space.

\end{abstract}

\newpage
\pagestyle{plain}
\pagenumbering{arabic}


\section{Introduction}

What is the most fundamental symmetry at the beginning of our universe? At that time, provided that 
several kinds of elementary particles existed, all particles would move very rapidly and could be effectively 
regarded as massless particles since not only the temperature of the universe is very high 
but also the vacuum energy is extremely huge. In such an effectively massless world, a scale symmetry 
would emerge automatically. It might be also natural to imagine that at the beginning of the universe, 
owing to strong quantum fluctuations, space-time itself fluctuates so violently that the continuous 
structure of the space-time does not make sense and therefore we have to consider somehow a discrete 
space-time which does not exhibit the properties of the manifold any longer. Even in such a discrete world, 
the scale symmetry might play a role since it can be formally defined by changing a scale at will at the location 
of each lattice site. 

From this perspective, we would like to consider the scale symmetry as the more fundamental symmetry rather than
general coordinate symmetry and the other symmetries around the beginning of the universe, and derive general relativity 
via its symmetry breakdown. Since we consider general relativity, it is reasonable to take account of not a global scale symmetry 
but a local scale symmetry, which we will call Weyl symmetry in what follows. Thus, the important problem is to start with a theory 
with Weyl symmetry and derive general relativity via some symmetry breakdown.

The Weyl symmetry in two space-time dimensions has played a pivotal role in string theory 
where it enabled that open and closed strings have a dual relation, 
and gauge theory and gravity are unified and at the same time the problem of ultraviolet divergences is automatically 
disappeared. However, in four space-time dimensions the Weyl symmetry has not played a role 
so far.\footnote{It is well known that we can construct a conformal gravity by taking the square term of conformal tensor 
in the action which is invariant under the Weyl transformation in four space-time dimensions, but this theory neither can avoid 
the problem of negative probabilities owing to the presence of four-derivative terms nor is renormalizable at least
perturbatively.}  In this article, we will fully utilize the Weyl symmetry in four dimensions and attempt to derive the Einstein-Hilbert action 
of general relativity through a symmetry breaking since we have neither the Weyl invariance nor the scale invariance 
in the present low-energy universe.

\section{A Weyl invariant theory}

It is physically plausible to suppose that the universe at the beginning was in a state of high energy and high temperature  
where a large vacuum energy triggers inflation of the universe. Under such a situation, it is of interest to conjecture that the starting
action might be made out of only a cosmological constant $\Lambda$:
\begin{eqnarray}
S_0 = \frac{\Lambda}{16 \pi G} \int d^4 x \, \sqrt{-g},
\label{Cosmo-action}  
\end{eqnarray}
where $G$ is Newton's constant and $g$ is a determinant of the metric tensor $g_{\mu\nu}$.\footnote{We follow the
notation and conventions of the MTW textbook \cite{MTW}.}

In this article, we assume that Weyl symmetry, or equivalently local scale symmetry, is the most fundamental symmetry 
at the beginning of the universe. However, the classical action (\ref{Cosmo-action}) is not invariant under the Weyl transformation. 
To remedy this issue, we introduce a scalar field $\phi$ and rewrite the action (\ref{Cosmo-action}) as
\begin{eqnarray}
S_c = \lambda \int d^4 x \, \sqrt{-g} \, \phi^4,
\label{Class-action}  
\end{eqnarray}
where $\lambda$ is a constant. It is then easy to prove that $S_c$ is invariant under the Weyl transformation:
\begin{eqnarray}
g_{\mu\nu} \rightarrow \Omega^2(x) g_{\mu\nu}, \qquad
\phi \rightarrow \Omega^{-1}(x) \phi,
\label{Weyl}  
\end{eqnarray}
where the transformation parameter $\Omega (x)$ depends on the space-time coordinates $x^\mu$.
Indeed, with a gauge choice $\phi = v$ such that $\lambda v^4 \equiv \frac{\Lambda}{16 \pi G}$, 
the action (\ref{Class-action}) becomes (\ref{Cosmo-action}), so the two actions are equivalent.

Since it is not classical mechanics but quantum mechanics that plays a role at the beginning of the universe, we must turn our attention 
to quantum field theory based on the classical action (\ref{Class-action}). This action is invariant under the Weyl transformation (\ref{Weyl})
and the general coordinate transformation (GCT), but since we regard the Weyl symmetry as the fundamental symmetry, we consider 
a BRST transformation only for the Weyl transformation, which is given by\footnote{As for the GCT, one can also introduce the
corresponding BRST transformation after introducing the BRST transformation for the Weyl transformation, if necessary.}  
\begin{eqnarray}
&{}& \delta_B g_{\mu\nu} = 2 c g_{\mu\nu}, \qquad
\delta_B \phi = - c \phi,  \qquad
\delta_B \bar c = i B,
\nonumber\\
&{}& \delta_B c = \delta_B  B = 0,
\label{BRST}  
\end{eqnarray}
where $c$ is the Faddeev-Popov (FP) ghost, $\bar c$ the FP antighost, and $B$ the Nakanishi-Lautrup
auxiliary field. This BRST transformation is obviously nilpotent, $\delta_B^2 = 0$ \cite{Kugo-Ojima, N-O-text}.

To fix the Weyl invariance, we have to pick up a gauge fixing condition, which was previously
taken to be $\phi = v$, by which the cosmological constant was generated. In the case at hand,
instead of the cosmological constant, we would like to generate the Einstein-Hilbert action.
For this purpose, we shall choose a gauge condition \cite{Oda-R}
\begin{eqnarray}
R = 0.
\label{R=0}  
\end{eqnarray}
One of the advantages of this gauge condition is that the scalar curvature transforms covariantly under 
a global scale transformation as seen shortly. Then, by following the standard recipe, the gauge fixing and FP ghost action 
can be constructed out of a BRST-exact form:
\begin{eqnarray}
S_{GF+FP} &=& - i \int d^4 x \, \delta_B ( \sqrt{-g} \, \bar c R )
\nonumber\\
&=& \int d^4 x \, \sqrt{-g} \left[ ( B + 2 i \bar c c ) R - 6 i \bar c \Box c \right],
\label{GF+FP-action}  
\end{eqnarray}
where we have used the BRST transformation (\ref{BRST}) and defined the d'Alembertian operator
$\Box$ for a scalar field $c$ by
\begin{eqnarray}
\Box c \equiv \frac{1}{\sqrt{-g}} \partial_\mu ( \sqrt{-g} g^{\mu\nu} \partial_\nu c ).
\label{d'Alem}  
\end{eqnarray}
Hence, the gauge fixed and BRST invariant quantum action takes the form:
\begin{eqnarray}
S_q &=& S_c + S_{GF+FP}
\nonumber\\
&=& \int d^4 x \, \sqrt{-g} \left[ ( B + 2 i \bar c c ) R + \lambda \phi^4 - 6 i \bar c \Box c \right].
\label{Quant-action1}  
\end{eqnarray}

We can further simplify this quantum action by redefining the auxiliary field $B$ as\footnote{The BRST
transformation for $\tilde B$ takes the form $\delta_B \tilde B = - 2 \tilde B c$.}  
\begin{eqnarray}
\tilde B \equiv B + 2 i \bar c c,
\label{Tilde-B}  
\end{eqnarray}
by which (\ref{Quant-action1}) can be rewritten into the form:
\begin{eqnarray}
S_q = \int d^4 x \, \sqrt{-g} \left( \tilde B R + \lambda \phi^4 - 6 i \bar c \Box c \right).
\label{Quant-action1-1}  
\end{eqnarray}
Furthermore, adding the following BRST-exact action $\tilde S_\alpha$ involving a gauge parameter $\alpha$
\begin{eqnarray}
\tilde S_\alpha = \int d^4 x \, \delta_B \left( - i \frac{1}{2} \alpha \sqrt{-g} \, \bar c \tilde B \right) 
= \int d^4 x \, \sqrt{-g} \, \frac{1}{2} \alpha \tilde B^2,
\label{Quant-action1-2}  
\end{eqnarray}
to $S_q$ and performing an integral over $\tilde B$, we can obtain an action 
\begin{eqnarray}
\tilde S_q = \int d^4 x \, \sqrt{-g} \left( - \frac{1}{2 \alpha} R^2 + \lambda \phi^4 - 6 i \bar c \Box c \right).
\label{Quant-action1-3}  
\end{eqnarray}

Two remarks are in order. First, the action (\ref{Quant-action1-3}) is invariant under a global scale
transformation since both (\ref{Quant-action1-1}) and (\ref{Quant-action1-2}) are so.
Here the scale transformation is defined by
\begin{eqnarray}
&{}& g_{\mu\nu} \rightarrow \Omega^2 g_{\mu\nu}, \qquad
\phi \rightarrow \Omega^{-1} \phi, \qquad
B \rightarrow \Omega^{-2} B,
\nonumber\\
&{}& c \rightarrow \Omega^{-1} c, \qquad
\bar c \rightarrow \Omega^{-1} \bar c, \qquad
\tilde B \rightarrow \Omega^{-2} \tilde B,
\label{Scale transf}  
\end{eqnarray}
where $\Omega$ is a constant. As mentioned before, it is easy to verify that under the scale transformation, 
the scalar curvature transforms covariantly as $R \rightarrow \Omega^{-2} R$. Second, this action represents that an $R^2$ gravity 
becomes a trivial theory if the ghost kinetic term is added to the action.

\section{Emergence of general relativity}

As long as we make use of the action (\ref{Quant-action1-2}) and proceed along the same line as above, we are forced to
reach the action (\ref{Quant-action1-3}) which is a rather trivial theory and has nothing to do with Einstein's gravity. 
In order to get over this problem, we need to take a different strategy. To do that, let us return to Eq. (\ref{Quant-action1-1}),
recall that this action has a global scale symmetry as a residual symmetry, and notice that it is a sort of scalar-tensor gravity \cite{Fujii} 
where the auxiliary field $\tilde B$ plays a role as a scalar field.

It is well known that in order that an action in the scalar-tensor gravity produces general relativity at low energies, it is necessary 
that the scale symmetry must be broken by some symmetry breaking mechanism \cite{Fujii}.\footnote{There is no trace anomaly 
since the classical action is just the cosmological term.} However, in the action (\ref{Quant-action1-1}) there are no suitable Higgs potentials 
for the scalar fields $\tilde B$ and $\phi$ which spontaneously break the scale symmetry.\footnote{For instance, the absence
of the Higgs potential for $\phi$ implies the VEV, $\langle \phi \rangle = 0$, but this VEV does not break the scale symmetry.}  
Thus it is reasonable to suppose that instead the strong gravitational interaction dynamics would drive the composite operator 
$\bar c c$ to ``condense into the vacuum'' such that $\langle \bar c c \rangle$ becomes nonzero since the gravitational
interaction could be strong at high energies around the Planck scale.\footnote{The auxiliary field $B$ in $\tilde B$ is proportional to
a BRST transformation of $\bar c$, so its VEV is vanishing, i.e. $\langle B \rangle = 0$.}  
To put it differently, it is expected that in the action (\ref{Quant-action1-1}), the vacuum expectation value of a composite operator 
$\bar c c$, $\langle \bar c c \rangle$ takes a finite value such that\footnote{It is reasonable that the Newton constant appears on the
right-hand of this equation since we assume that the condensation occurs around the Planck scale and there are no the other
scales owing to the scale invariance in the action (\ref{Quant-action1-1}).} 
\begin{eqnarray}
2 i \langle \bar c c \rangle = \frac{1}{16 \pi G}.
\label{VEV}  
\end{eqnarray}

Of course, the product of two Heisenberg operators at the same space-time point is in general not well defined,
so we have to start by defining the local product neatly by introducing some regularization and
renormalization procedures.
This study is beyond the scope of this article and we simply assume that the product of operators 
at the same space-time point is defined by a limiting procedure: 
\begin{eqnarray}
\langle \bar c c \rangle \equiv \lim_{x \to y} \langle \bar c(x) c(y) \rangle.
\label{Product}  
\end{eqnarray}
Assuming the uniform convergence, we can obtain 
\begin{eqnarray}
\langle \bar c \Box c \rangle &=& \lim_{x \to y} \langle \bar c(x) \frac{1}{\sqrt{-g}} \partial_\mu ( \sqrt{-g} g^{\mu\nu}
\partial_\nu c(y) ) \rangle
\nonumber\\
&=& \lim_{x \to y} \frac{1}{\sqrt{-g(y)}} \partial_\mu^y ( \sqrt{-g(y)} g^{\mu\nu}(y)
\partial_\nu^y ) \langle \bar c(x) c(y) \rangle
\nonumber\\
&=& 0.
\label{Product2}  
\end{eqnarray}

It is of interest that this result can be also obtained from a completely different argument. Recall that
the first and third terms on the right-hand side (RHS) of Eq. (\ref{Quant-action1-3}) come from a BRST-exact form, so we can regard
a VEV of the third term as a negative VEV of the first term
\begin{eqnarray}
- 6 i \langle \bar c \Box c \rangle = \frac{1}{2 \alpha} \langle R^2 \rangle.
\label{Two-VEV}  
\end{eqnarray}
Then, the RHS of this equation approaches to zero if the gauge parameter $\alpha$ is chosen to be much larger than $\langle R^2 \rangle$. 
  
Under such a situation, the action reduces to the form:
\begin{eqnarray}
S_q = \int d^4 x \, \sqrt{-g} \left( \frac{1}{16 \pi G} R + \lambda \phi^4 \right).
\label{EH}  
\end{eqnarray}
This is nothing but the Einstein-Hilbert action with a field-dependent cosmological term.
In this sense, general relativity (plus a field-dependent cosmological term) is emerged from a cosmological constant via ghost 
condensation.\footnote{Our ghost condensation mechanism is different from that of Refs. \cite{Arkani, Oda-M}.}
Let us note that our symmetry breaking mechanism is an example of dynamical symmetry breaking.

It is worthwhile to consider why general relativity is induced by the ghost condensation of
the FP ghosts associated with the Weyl symmetry. In the classical theory, there are no dynamical
degrees of freedom since the classical action is equivalent to the cosmological constant.
After the BRST quantization, two fermionic degrees of freedom associated with the FP ghost and
antighost are introduced in a quantum theory, but the ghost condensation decreases these two degrees
of freedom, so as a result two bosonic degrees of freedom, which correspond to two physical 
degrees of freedom associated with the graviton, are created.

Incidentally, we can understand why the gauge condition (\ref{R=0}) takes a special position among many of gauge conditions
for the Weyl symmetry. This is because it is only this gauge condition that leaves a global scale symmetry as a residual
symmetry. For instance, if we chose the gauge condition $R + R^2 = 0$, we could not get the scale symmetry
after the gauge fixing. Of course, it is possible to obtain a scale invariant gauge condition by modifying this gauge
condition as $R + \frac{1}{\phi^2} R^2 = 0$, but then we would have a non-polynomial interaction or a singularity at
$\phi = 0$. Thus, in the present formalism, a gravitational theory described by a Lagrangian density $\sqrt{-g} (R + R^2)$, 
or even the more general $f(R)$ gravity cannot be derived and only Einstein's general relativity can be induced.  

This fact is also consistent with the following observation with respect to the number of physical degrees of freedom:
We have now two remaining bosonic degrees of freedom which precisely coincide with the number of those of
the graviton in Einstein's general relativity. On the other hand, in the $f(R)$ gravity, in addition to the graviton
there exists one scalar field called ``scalaron'' \cite{Oda-f}, so we cannot generate the $f(R)$ gravity within the framework
of the present formalism.  

It is also possible to interpret our symmetry breaking mechanism as a bound state problem. 
First, let us take a BRST transformation of the composite operator $\bar c c$ which is given 
by
\begin{eqnarray}
\delta_B ( \bar c c ) = i B c \equiv \Psi.
\label{Comp-BRST}  
\end{eqnarray}
Here we introduce $\bar \Psi$ as a conjugate operator to $\Psi$ defined by
\begin{eqnarray}
\bar \Psi \equiv i B \bar c.
\label{Conj-op}  
\end{eqnarray}
The BRST transformation of this operator yields
\begin{eqnarray}
\delta_B \bar \Psi = - B^2 \equiv \Gamma.
\label{Conj-BRST}  
\end{eqnarray}
These equations represent that when the bound state $\bar c c$ is formed, there should be bound states in the channels of $B c, B \bar c$ 
and $B^2$ as well. Then, it is easy to see that the operators $\{ \bar c c, \Psi, \bar \Psi, \Gamma \}$ constitute
a BRST quartet and their asymptotic states therefore belong to unphysical Hilbert space \cite{Kugo-Ojima}. 
In particular, this fact implies that an asymptotic state of the bound state $\bar c c$ made from the FP ghosts $c$ and $\bar c$ is ``confined'' to 
the unphysical sector and cannot be observed as in those of the FP ghosts $c$ and $\bar c$ themselves. In other words, the VEV
of the bound state $\bar c c$ takes the definite value in Eq. (\ref{VEV}) but the bound state has no corresponding asymptotic state, so
the associated Nambu-Goldstone boson is not left behind after symmetry breakdown of the scale symmetry.

\section{Conclusions}

In string theory and conformal field theory in two dimensions, Weyl symmetry, or equivalently local scale symmetry has 
played an important role, but thus far it has not been utilized in four space-time dimensions owing to the existence 
of the massive ghost violating the unitarity although many of attempts for overcoming this problem
have recently been proposed \cite{Luca}. Since we have no Weyl symmetry and global scale symmetry 
in our present world, they must be broken by some ingenious mechanism. One of the most interesting mechanism 
of the symmetry breaking would be certainly a spontaneous symmetry breaking where a scalar field acquires 
a nonvanishing vacuum expectation value, but both the Weyl and global scale symmetries do not allow 
the presence of coupling constants with mass dimension, thereby making it impossible to write down the Higgs potential 
of the Mexican-hat type.   

It is therefore desirable to develop an alternative mechanism of symmetry breaking for both local and global
scale symmetries. In this respect, quantum chromodynamics (QCD) gives us useful information. In QCD,
there is no elementary scalar field to acquire a vacuum expectation value so the composite scalar
fields such as $\bar u u$ and $\bar d d$ where $u, d$ are up quark and down quark, respectively, condense 
into the vacuum and develop nonvanishing values. In the case at hand, the situation is similar to that of QCD,
but the composite operator $\bar c c$ belongs to unphysical Hilbert sector as in $\bar c, c$ and leaves
nothing behind after the symmetry breaking whereas the composite operators $\bar u u$ and $\bar d d$ belong to 
physical Hilbert space, thereby creating the observed hadrons.  This ``confinement'' of the bound state $\bar c c$
of the FP ghosts to unphysical Hilbert space is the key ingredient for our symmetry breaking mechanism.

We should comment on an important remaining problem of the formalism at hand. Within the present framework,
it seems to be rather difficult to prove that the composite operator $\bar c c$ has the nonvanishing vacuum 
expectation value since the physics behind this problem is in the strong coupling region so we cannot make use of 
perturbation theory and need a non-perturbative method. However, we have so far observed experimentally 
only the Higgs particle as a physical scalar particle (which appears after the spontaneous symmetry breakdown 
of the gauge symmetry) and it is usually expected that in any case the scalar field would condense into the vacuum, 
so it might be possible to assume that the scalar composite operator $\bar c c$ could also develop a nonzero vacuum 
expectation around the Planck energy owing to the strong gravitational interaction. Then, it is also plausible to suppose that 
the composite operator $\bar c c$ constitutes a bound state. Once such a bound state is formed, the composite operator 
$\bar c c$ is not invariant under the BRST transformation so it belongs to unphysical sector by the BRST quartet mechanism. 
In any case, to show these ideas explicitly, there are a lot of works to be done in future.

Finally, it is useful to mention a relation between the present theory and pregeometry (or induced gravity) \cite{Sakharov,
Akama, Amati, Adler, Floreanini, Oda-A1, Oda-A2}. In a scenario of pregeometry, gravity is not fundamental but induced from one-loop 
effects of matter fields via the ultraviolet cutoff at the Panck scale whereas in the present theory, gravity is not fundamental either 
but induced from the cosmological constant via the condensation of FP ghosts. Although the both theories are quite different,
they share one important point: Even if Einstein gravity is not there at classical level, it will automatically be generated at
quantum level.


\end{document}